\documentclass[%
 aip,
 amsmath,amssymb,
 reprint,%
]{revtex4-1}

\usepackage{graphicx}% Include figure files
\usepackage{dcolumn}% Align table columns on decimal point
\usepackage{bm}% bold math
\usepackage[usenames,dvipsnames]{color}
\usepackage[utf8]{inputenc}
\usepackage[T1]{fontenc}
\usepackage{amsmath,amssymb}
\usepackage[mathcal,mathscr]{eucal}
\usepackage{times}
\usepackage{lmodern}
\usepackage[Euler]{upgreek}
\usepackage{enumerate}
\usepackage{textcomp}
\usepackage{stmaryrd}
\usepackage[colorlinks=true, allcolors=blue]{hyperref}

\DeclareUnicodeCharacter{2212}{-}
\DeclareUnicodeCharacter{0001}{-}
\definecolor{gray}{RGB}{191,191,191}
\definecolor{darkgreen}{RGB}{0,191,0}
\definecolor{purple}{RGB}{191,127,127}
\definecolor{orange}{RGB}{255, 165, 0}

\begin{document}

\preprint{AIP/123-QED}

\title[Surface reconstruction of tetragonal methylammonium lead triiodide]{Surface reconstruction of tetragonal methylammonium lead triiodide}
% Force line breaks with \\

\author{Azimatu Seidu}
\email{azimatu.seidu@aalto.fi}
\affiliation{Department of Applied Physics, Aalto University, FI-00076 AALTO, Finland}
\author{Marc Dvorak}
\affiliation{Department of Applied Physics, Aalto University, FI-00076 AALTO, Finland}
\author{Jari J\"arvi}
\affiliation{Department of Applied Physics, Aalto University, FI-00076 AALTO, Finland}
\author{Patrick Rinke}
\affiliation{Department of Applied Physics, Aalto University, FI-00076 AALTO, Finland}
\author{Jingrui Li}
\affiliation{Electronic Materials Research Laboratory, Key Laboratory of the Ministry of Education \& International Center for Dielectric Research, School of Electronic Science and Engineering, Xi'an Jiaotong University, Xi'an 710049, China}

\date{\today}% It is always \today, today,

\begin{abstract}
We present a detailed first-principles analysis of the (001) surface of methylammonium lead triiodide (MAPbI$_3$). With density-functional theory we investigate the atomic and electronic structure of the tetragonal ($I4cm$) phase of MAPbI$_3$. We analysed surfaces models with MAI- (MAI-T) and PbI$_2$\,-terminations (PbI$_2$\,-T). For both terminations, we studied the clean-surface and a series of surface reconstructions. We find that the clean MAI-T model is more stable than its PbI$_2$\,-T  counterpart. For the MAI termination, reconstructions with added or removed units of nonpolar MAI and PbI$_2$ are most stable. The corresponding band structures reveal surface states originating from the conduction band. Despite the presence of such additional surface states, our stable reconstructed surface models do not introduce new states within the band gap.
\end{abstract}
\maketitle

%\begin{quotation}
%The ``lead paragraph'' is encapsulated with the \LaTeX\ 
%\verb+quotation+ environment and is formatted as a single paragraph before the first section heading. 
%(The \verb+quotation+ environment reverts to its usual meaning after the first sectioning command.) 
%Note that numbered references are allowed in the lead paragraph.
%
%The lead paragraph will only be found in an article being prepared for the journal \textit{Chaos}.
%\end{quotation}

\section{Introduction}\label{intro}

Perovskite solar cells (PSCs) have attracted immense attention within the photovoltaic community due to their rapidly rising power conversion efficiency (PCE): it reached 25.5\% \cite{NRELchart19} only nine years after the invention of the state-of-the-art PSC architecture in 2012 (PCE $\sim\!$10\%) \cite{kim,lee}. The hybrid (organic-inorganic) halide perovskite (HP) methylammonium (MA) lead triiodide ($\text{CH}_3^{}\text{NH}_3^{}\text{PbI}_3^{}$ or $\text{MAPbI}_3^{}$) has been the most common PSC photoabsorber for a long time, and it is still a major focus of both experimental and theoretical studies, along with the rising isostructural material based on formamidinium (FA). HPs have also received significant recognition in luminescence and light detection \cite{Das2020, Giuia2021, Huang14, Lin15, Cho15, Li2020}.

To advance HPs for eventual use in large-scale commercial applications, further efforts in fundamental research are still necessary to enable materials and device engineering. Researching surface passivation is critical in this regard since defects at perovskite surfaces and grain boundaries are centers of nonradiative recombination, which is a major inhibitor to further PCE improvement \cite{Yin14, Steirer2016, Walsh15, Kim14, Ball16, Wu2015,  Long2016}. Additionally, organic components in hybrid HPs suffer from rapid degradation when exposed to moisture, heat, and oxygen \cite{NiuG14,NiuG15,HuangJ17,KimGH17,Mesquita18, LiF18}, the effects of which can be reduced with proper surface passivation. To enhance the stability of HPs and enable large-scale applications, measures need to be taken to mitigate their instability with minimal compromise to PCE. Several proposals have addressed this challenge. Notable approaches include surface passivation via organic long-chain ligands \cite{SchmidtL14, SoranyelG15,Dong19}, dimensionality reduction of perovskite active materials \cite{Quan16,Dou17,Ran18a,WangZ18,LiuC18,Ran19}, protective coating with inorganic semiconductors or insulators \cite{Matteocci16,Cheacharoen18a,Cheacharoen18b,Seidu19}, and A-site  substitution with smaller monovalent ions \cite{noh,Yi16,ZhouY16,Tan17,Ciccioli18,Gao18,Baena17,Ganose17}.

The application of these proposed solutions requires an understanding of the surface properties and possible surface reconstructions of HPs. This includes several aspects, such as morphology control during the growth of the HP thin films, HP-interlayer interface engineering, and the passivation of intrinsic defects at the interfaces and grain boundaries. A comprehensive understanding of the atomic and electronic structure of $\text{MAPbI}_3^{}$ surfaces would advance the development of this class of novel materials and their applications. The surfaces of MA- \cite{Haruyama14,Haruyama16, Akbari17, Zhang2016} and FA-based \cite{Wang2020, Xue2018, Fu2017} perovskites have been investigated theoretically and experimentally \cite{Jiang19, Chen18, Cho18, Saliba16a, Saliba16b, Saliba16c, Qiu2020, He2020, Das2020}, but understanding of the non-pristine surfaces is still lacking. Most of the  MA-perovskite surface studies are focused on the stability of the two main terminations of HP $(001)$ surfaces: MAI- and PbI$_2$\,-terminated (shortened as MAI-T and PbI$_2$\,-T hereafter, respectively) with little to no consideration of possible surface reconstructions.

In our previous work, we investigated the atomic and electronic structure of $(001)$ surfaces of cesium lead triiodide (CsPbI$_3$) using first-principles density functional theory (DFT) calculations and surface-phase-diagram (SPD) analysis \cite{Seidu2021}. For both cubic ($\upalpha$) and orthorhombic ($\upgamma$) phases, we found that the CsI-termination is more stable than PbI$_2$\,-termination, and the former class features a series of stable surface reconstructions with added or removed valence-neutral CsI and PbI$_2$ units. Our previous study established a systematic method for understanding stable surface reconstructions with a representative HP and motivates the present work in regard to both materials engineering and theoretical methodology.
% The symmetric-slab-model approach effectively prevented the possible artefact of new states in the band gap.

In this work, we present a comprehensive DFT study of the $(001)$ surface of the room-temperature tetragonal phase of the more popular HP, $\text{MAPbI}_3^{}$\,. Haruyama et~al. have carried out preliminary studies for this surface and identified some stable surface terminations dependent on growth conditions, with some of them beyond the regular ``clean surface'' models \cite{Haruyama14,Haruyama16}. Nevertheless, an extensive exploration of surface terminations and reconstructions with the addition or removal of constituent elements $\text{CH}_3^{}\text{NH}_2^{}$ (MeNH$_2$), $\text{Pb}$, $\text{I}$, and their complexes is lacking. We aim to establish such a systematic theoretical description by means of DFT, \textit{ab initio} thermodynamics \cite{karsten, reuter, abinitiothermodynamics}, and SPD analysis.

It is worth noting that the unique charge state of the organic MA cation introduces additional complexity into this DFT study compared to our work on CsPbI$_3$. Simply separating MAPbI$_3$ into its constituents MA, Pb, and I in a way similar to the decomposition of CsPbI$_3$ into Cs, Pb, and I$_2$ is not thermodynamically sensible. The charge-neutral $\text{CH}_3^{}\text{NH}_3^{}$ radical is not stable on its own and far less suitable as a thermodynamic reference system than Cs is for CsPbI$_3$. We therefore use the neutral  $\text{CH}_3^{}{\text{NH}_2^{}}$ ($\text{MeNH}_2^{}$) and $\text{H}_2^{}$ molecules in this work. Similar to Ref.~\onlinecite{Seidu2021}, we will classify the thermodynamic stability of considered MAPbI$_3$ surfaces for different growth conditions and analyze their electronic structure.

The remainder of this paper is organized as follows. In Sec.~\ref{methods}, we briefly outline the computational details of our DFT calculations  and summarize the thermodynamic constraints for the growth of bulk MAPbI$_3$, as well as the MAI-T and PbI$_2$\,-T surfaces. In Sec.~\ref{results}, we first analyze the stability of the clean-surface models (MAI-T and Pb$_2$\,-T) and the reconstructed models with missing- and add-atoms and complexes. We then discuss the impact of surface reconstruction on both the atomic and electronic structure. Finally, we conclude with a summary in Sec.~\ref{conclusion}.

\section{Computational details}\label{methods}
All DFT calculations were performed using the Perdew-Burke-Ernzerhof exchange-correlation functional for solids (PBEsol) \cite{perdew08} implemented in the all-electron numeric-atom-centered orbital code \textsc{fhi-aims} \cite{Blum09,Havu2009,Levchenko/etal:2015}. We chose PBEsol because it describes the lattice constants of $\text{MAPbI}_3^{}$ well at moderate computational cost \cite{Yang17, Bokdam17}. In our previous study on  CsPbI$_3$ surfaces \cite{Seidu2021}, we also tested the PBE functional, but found only negligible changes in the surface phase diagram. We expect the same to be true for MAPbI$_3$. Scalar relativistic effects were included by means of the zeroth-order regular approximation \cite{vanlenthe}. As with the PBE test, the inclusion of full spin-orbit coupling did not affect the conclusions of our CsPbI$_3$ study \cite{Seidu2021} and we expect the same for MAPbI$_3$.  Standard \textsc{fhi-aims} tier-2 basis sets were used in combination with $\Gamma$-centered $4\times4\times4$ (bulk) and $4\times4\times1$ (surfaces) $k$-point meshes. The bulk structures were optimized with the analytical stress tensor \cite{knuth} until forces were below $5.0\times10^{-3}~\text{eV}\cdot\text{\AA}^{-1}$. For the surface slab models, we fixed the lattice constants and all atomic positions except for atoms in the top and bottom MAPbI$_3$ units (the surface atoms). A surface-dipole correction \cite{Neugebauer92} was applied in all surface calculations.

In the interest of open science \cite{Himanen2019}, we made all relevant calculations included in this work available on the Novel Materials Discovery (NOMAD) repository \cite{Note-NOMAD}.

\subsection{Structural optimization}

\subsubsection{Bulk and surface structures}
\label{sec:bulkstructure}

\begin{figure}[!htp]
\includegraphics[width=\linewidth]{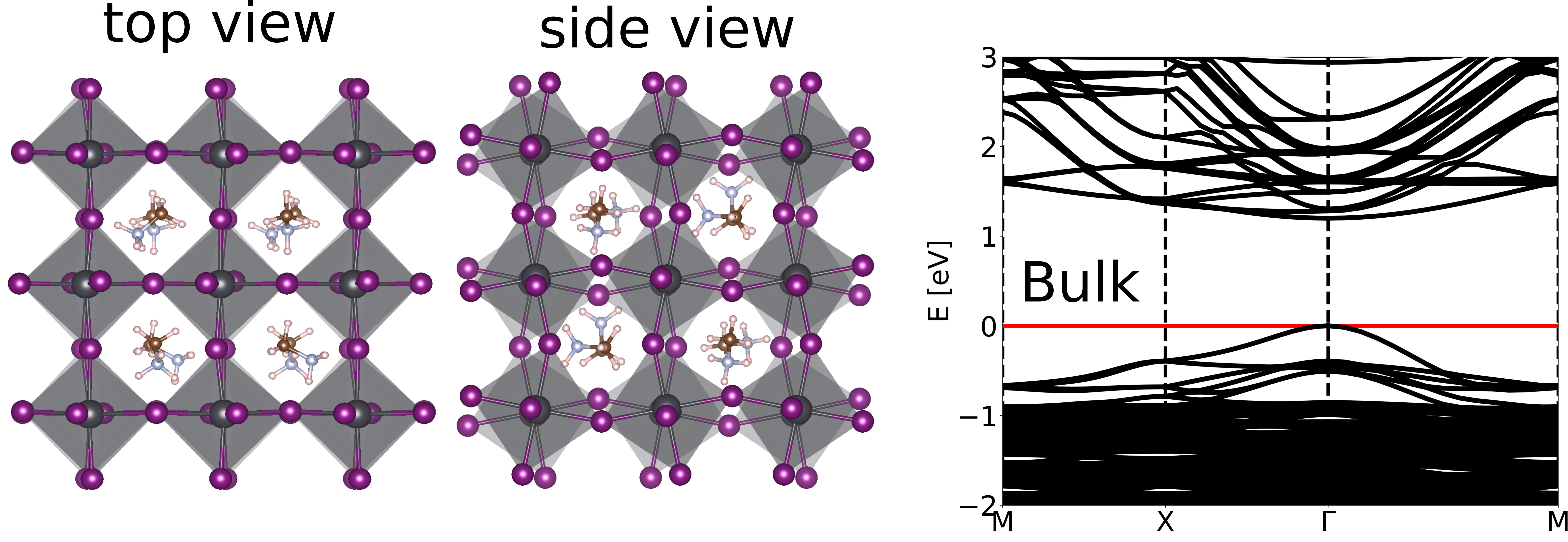}\vspace{-1.0em}
\caption{Bulk geometry and band structure of $I4cm$ phase of $\text{CH}_3^{}\text{NH}_3^{}\text{Pb}\text{I}_3^{}$ in the $2\times2\times2$ supercell model. $\text{C}$, $\text{H}$, $\text{N}$, $\text{Pb}$, and $\text{I}$ are colored in brown, light gray, light blue, black, and purple, respectively. The PbI$_6$ octahedra are colored in dark gray. The valence band maximum is set to zero and depicted by the red line in the band structure plot.}\label{bulk}
\end{figure}

As experimentally reported, the tetragonal phase of $\text{MAPbI}_3^{}$ is stable from $\sim\!160$ to $\sim\!330~\text{K}$, including room temperature \cite{Stoumpos13,Baikie13}. The structure belongs to the polar space group $I4cm$ (No.~108) as a result of its intrinsic polarization along the principal axis \cite{Stoumpos13}. Considering several possible disordered MA alignments \cite{Lahnsteiner16,li2018}, we constructed a series of $2\times2\times2$ supercells with different MA orientations and optimized their structures with DFT. We then take the structure with the lowest energy. The lattice parameters of this structure (Fig.~\ref{bulk}) are $a\!=\!b\!=\!12.40$~{\AA}, $c\!=\!12.68$~{\AA}. Figure~\ref{bulk} displays some disorder and an overall vertical (downward in the side view, i.e., $[00\bar{1}]$) net dipole, which is formed by the $\text{C--N}$ dipoles of the MA cations. The horizontal, i.e., $(001)$, component of the overall dipole moment within the model nearly vanishes.

\begin{figure}[!htp]
\includegraphics[width=\linewidth]{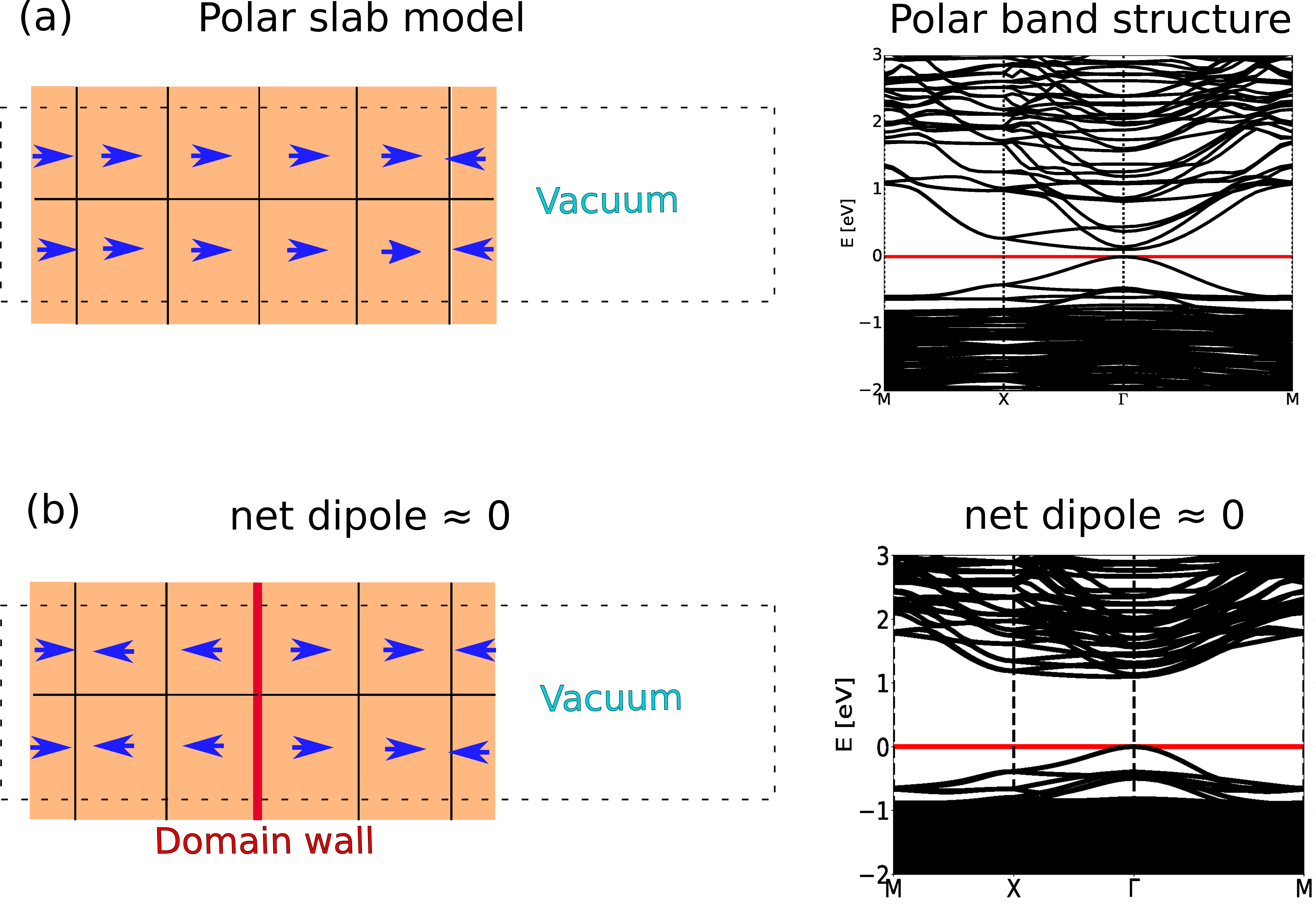}\vspace{-1.0em}
        \caption{Construction of nonpolar MAI-T slab model, with the $[001]$ components of MA dipoles represented by blue arrows. (a) A polar slab model results from simple repetition of bulk unit cells. (b) By mirroring bulk structures around a central domain wall, the overall dipole vanishes. Also shown is how the surface dipoles on both sides of the slab reorient themselves during relaxation and point inwards as a result of hydrogen bonding with surface I ions. The band structure of each model is given in the right column. } \label{slab_mod}
\end{figure}

In this work, we focus on the $(001)$ surfaces, which are the major facet of HPs \cite{Haruyama14,Haruyama16,Schulz19} and the most relevant surfaces of $\text{MAPbI}_3^{}$\,. Due to the polar bulk structure, it is not possible to build a surface supercell by repeating several bulk layers along the $[001]$ direction as this would result in a polar surface model (see Figure \ref{slab_mod}a left). Such a model will induce artefacts into the calculated properties of the system such as the unphysical removal of band degeneracies and reduction of the band gap (Fig.~\ref{slab_mod}a right) and would ultimately lead to a polar catastrophe, in which the valence band at one end of the slab lie higher in energy than the conduction bands at the other end.  %\PRc{should we mention here that previous work done by others was suffering from these problems? \AS{I haven't seen any literature that talks about this specifically. Maybe Jingrui might have seen any.}} \JL{I agree with Patrick. Didn't we mention in our meeting that the Haruyama 2014 paper did mention ``something like that'', although not quite clearly and not quite the same? Search ``dipole'' or ``symmetr'' in their paper to get more information. We can simply say that similar approach was successfully applied in previous study/-ies, and cite this paper (and check whether Roiati Nano Lett. 14, 2168 (2014) is also related). \AS{Done}\PRc{Did previous studies address the polar catastrophe or did they not? I thought they didn't. In any case, the ``Similar approaches sentences'' needs to be moved to the end of the next paragraph, because in this paragraph we are only raising the problem, but we are not addressing it.}}

To circumvent these artefacts, we constructed a symmetric slab model by introducing a ``domain wall'' in the slab. As sketched in Figure~\ref{slab_mod}b left, such a domain wall is a $\text{PbI}_2^{}$\,-containing $(001)$ plane located at the center of the slab. The atomic structures on opposite sides of the domain wall is mirrored with respect to this plane, so that the $[001]$ components of the MA dipole moments on opposite sides cancel each other, giving rise to a nearly vanishing overall dipole moment. As a result, the polar artefacts vanish and the  surface band structure (Fig.~\ref{slab_mod}b right) exhibits a proper band gap and the right degeneracies. Similar approaches have been successfully employed in previous studies \cite{Haruyama14, Roiati2014}.

With the approach illustrated in Fig.~\ref{slab_mod}b, we constructed symmetric clean surface models in a way similar to our previous work for CsPbI$_3^{}$\,. Specifically, the MAI-T surface model consists of 6 MAI and 5 PbI$_2$ layers alternately stacked along the $[001]$ direction. Similarly, the PbI$_2$\,-T surface model has 7 PbI$_2$ and 6 MAI alternating layers. By inserting a $40~\text{\AA}$-thick vacuum layer to separate neighboring slabs along $[001]$ and including surface-dipole correction \cite{Neugebauer92} in the DFT calculations, we minimized the interaction between neighboring slabs. 

\begin{figure}[!htp]
\centering
\includegraphics[width=0.7\linewidth]{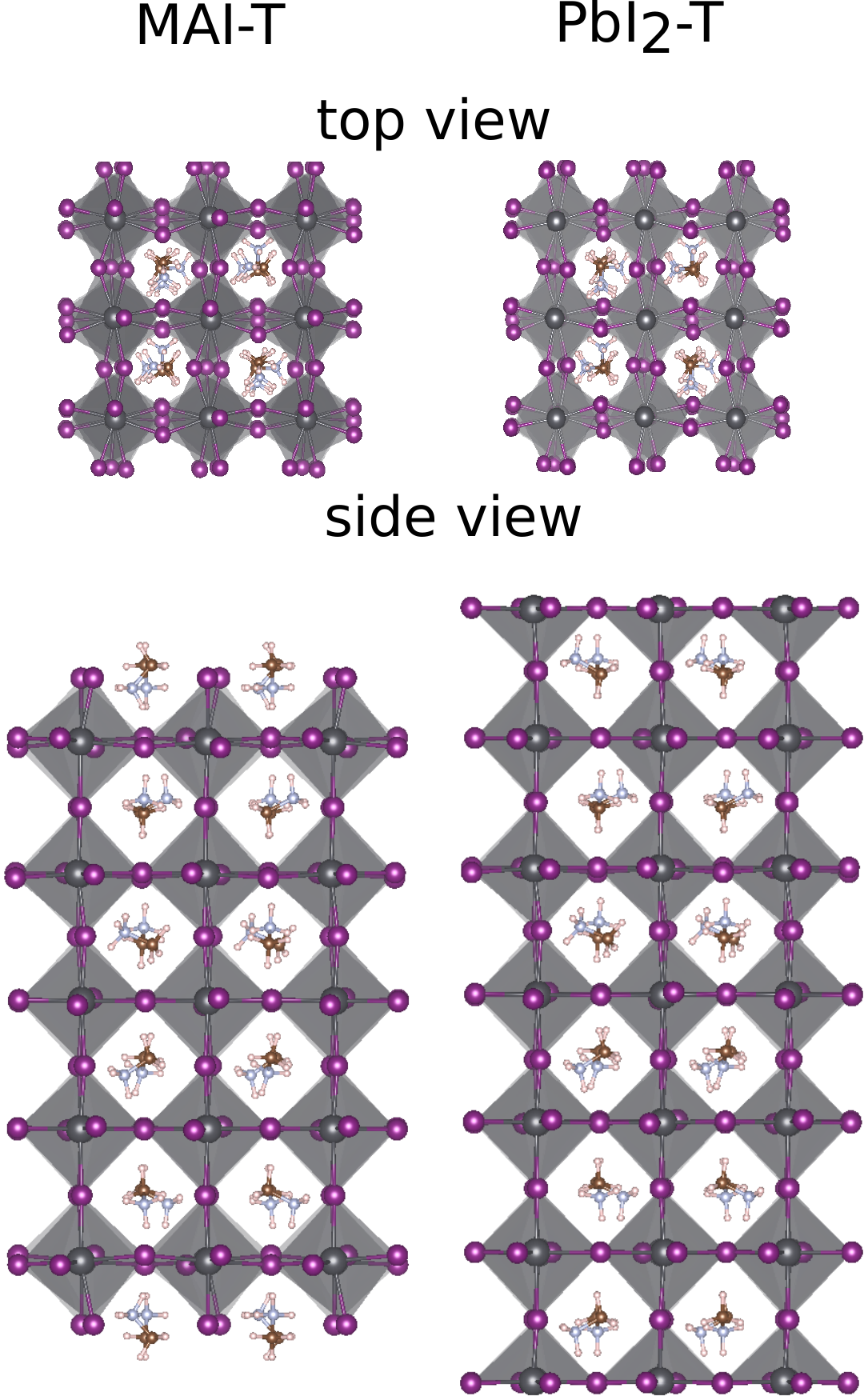}\vspace{-1.0em}
\caption{Relaxed $\text{MAI}$-T and $\text{PbI}_2^{}$\,-T clean-surface models. Depicted on the left is the MAI-T and on the right the $\text{PbI}_2^{}$\,-T termination.}\label{surfaces}
\end{figure}

Figure~\ref{surfaces} depicts the optimized structures of both clean MAI-T and PbI$_2^{}$\,-T surfaces. The top views of both phases show a similar in-plane tilting pattern of PbI$_6^{}$ octahedra and in-plane alignment of MA dipoles as in the bulk. The side views demonstrate that the mirror symmetry of both slab models with respect to the domain wall is maintained after geometry optimization. We note that in MAI-T, MA dipoles at both top and bottom surfaces point inwards (sketched in Fig.~\ref{slab_mod}) as a result of hydrogen bonding with the surface I ions.

We studied various add- and missing-atom surface models based on both MAI-T and $\text{PbI}_2^{}$\,-T clean surfaces. All add-atom models (i$_{\text{X}}^{}$) were constructed by adding the atoms or atom-complexes X to the surface, while for missing-atom models (v$_{\text{X}}^{}$), atoms or complexes X were removed from the topmost X-containing layers. For $\text{MAI}$-T surfaces as an example, v$_{\text{MeNH}_2^{}}$\,, v$_{\text{H}}$\,, v$_{\text{\text{MA}}}$\,, v$_{\text{I}}$\,, and v$_{\text{MAI}}$ were constructed by removing atoms from the topmost MAI layer, while v$_{\text{Pb}}$\, and v$_{\text{PbI}_2^{}}$ indicate the removal of atoms from the PbI$_2$ layer below the topmost MAI layer. For models with double missing- or add-atoms (i.e., v$_{2\text{X}}$ or i$_{2\text{X}}$), we considered both line and diagonal modes that correspond to the reconstruction units distributed along the $[100]$ or $[110]$ directions, respectively. Only the more stable model will be presented and discussed in Sec.~\ref{results}. For instance, we find the line modes to be more stable in both v$_{2\text{MAI}}$ and i$_{2\text{PbI}_2^{}}$.

\begin{table}[!htp]
\caption{Reconstructed MAI-T and $\text{PbI}_2^{}$\,-T surface models of tetragonal MAPbI$_3$ considered in this work.} \label{recon_struc}
\begin{tabular}{ll@{\hspace{2em}}ll} \hline\hline
\multicolumn{2}{c@{\hspace{2em}}}{MAI-T} & \multicolumn{2}{c}{PbI$_2^{}$\,-T} \\ \hline
v$_{\text{MeNH}_2^{}}$ & i$_{\text{MeNH}_2^{}}$ & v$_{\text{MeNH}_2^{}}$ & i$_{\text{MeNH}_2^{}}$ \\
v$_{2\text{MeNH}_2^{}}$ & i$_{2\text{MeNH}_2^{}}$ & v$_{2\text{MeNH}_2^{}}$ & i$_{2\text{MeNH}_2^{}}$ \\
v$_{\text{MA}}$ & i$_{\text{MA}}$ & v$_{\text{MA}}$ & i$_{\text{MA}}$ \\
v$_{2\text{MA}}$ & i$_{2\text{MA}}$ & v$_{2\text{MA}}$ & i$_{2\text{MA}}$ \\
v$_{4\text{MA}}$ & i$_{\text{Pb}}$ & v$_{\text{Pb}}$ & i$_{\text{Pb}}$ \\
v$_{\text{Pb}}$ & i$_{2\text{Pb}}$ & v$_{2\text{Pb}}$ & i$_{2\text{Pb}}$ \\
v$_{2\text{Pb}}$ & i$_{\text{I}}$ & v$_{4\text{Pb}}$ & i$_{\text{I}}$ \\
v$_{\text{I}}$ & i$_{2\text{I}}$ & v$_{\text{I}}$ & i$_{2\text{I}}$ \\
v$_{2\text{I}}$ & v$_{\text{H}}$ & v$_{2\text{I}}$ & i$_{\text{H}}$ \\
v$_{\text{H}}$ & i$_{2\text{H}}$ & v$_{\text{H}}$ & i$_{2\text{H}}$ \\
v$_{2\text{H}}$ & i$_{4\text{H}}$ & v$_{2\text{H}}$ & i$_{4\text{H}}$ \\
v$_{4\text{H}}$ & i$_{\text{MAI}}$ & v$_{4\text{H}}$ & i$_{\text{MAI}}$ \\
v$_{\text{MAI}}$ & i$_{2\text{MAI}}$ & v$_{\text{MAI}}$ & i$_{2\text{MAI}}$ \\
v$_{2\text{MAI}}$ & i$_{4\text{MAI}}$ & v$_{2\text{MAI}}$ & i$_{4\text{MAI}}$ \\
v$_{4\text{MAI}}$ & i$_{\text{PbI}_2^{}}$ & v$_{\text{PbI}_2^{}}$ & i$_{\text{PbI}_2^{}}$ \\
v$_{\text{PbI}_2^{}}$ & i$_{2\text{PbI}_2^{}}$ & v$_{2\text{PbI}_2^{}}$ & i$_{2\text{PbI}_2^{}}$ \\
v$_{2\text{PbI}_2^{}}$ & i$_{4\text{PbI}_2^{}}$ & v$_{4\text{PbI}_2^{}}$ & i$_{4\text{PbI}_2^{}}$ \\ \hline\hline
\end{tabular}
\end{table}

\subsection{Thermodynamic constraints for stable \texorpdfstring{$\text{MAPbI}_3^{} \;${}} bbulk and surfaces}\label{therm_stab}

We applied the grand potential analysis to investigate the stability of a variety of different surface reconstructions. Neglecting finite temperature contributions, the grand potential ($\varOmega$) is
\begin{equation}
    \varOmega = \Delta H - \sum_i x_i \Delta \mu_i = E - \sum_i x_i \mu_i^{\minuso} - \sum_i x_i \Delta \mu_i \, \label{grandpot.}.
\end{equation}
Here, $\Delta H$ indicates the standard formation energy of the model system, $E$ the total energy, $\mu_i^{\minuso}$ the chemical potential of species $i$ in its most stable form, $x_i$ the number of atoms of this species in the structure, and $\Delta \mu_i$ the change in the chemical potential away from its value in the element's most stable phase, $\mu_i^{\minuso}$\,. $\Delta \mu_i$ represents the control of experimental growth conditions and is both a meaningful and convenient parameter to vary in phase diagrams. The relative stability between two structures is determined by comparing their grand potentials, with the structure lower in grand potential considered more stable. Details of the grand potential analysis are described in our previous work on surface reconstruction of CsPbI$_3$ \cite{Seidu2021}.

We first consider conditions for stable $\text{MAPbI}_3^{}$ in the bulk. In order to avoid the formation of elemental Pb and I, molecular MA (as a whole instead of elemental C, N, and H for simplicity), as well as bulk MAI and $\text{PbI}_2^{}$\,, the region of the phase diagram for stable $\text{MAPbI}_3^{}$ is determined by the inequalities,
\begin{equation}
\begin{split}
\Delta H (\text{MAPbI}_3^{}) &\leqslant \Delta\mu_{\text{MA}} \leqslant 0 \, , \\
\Delta H (\text{MAPbI}_3^{}) &\leqslant \Delta\mu_{\text{Pb}} \leqslant 0 \, , \\
\Delta H (\text{MAPbI}_3^{}) &\leqslant 3 \Delta\mu_{\text{I}} \leqslant 0 \, ;
\end{split} \nonumber
\end{equation}
and
\begin{equation}
\begin{split}
\Delta H (\text{MAPbI}_3^{}) &\leqslant \Delta\mu_{\text{MA}} + \Delta\mu_{\text{Pb}} + 3\Delta\mu_{\text{I}} \, , \\
\Delta\mu_{\text{MA}} + \Delta\mu_{\text{I}} &\leqslant \Delta H (\text{MAI}) \, , \\
\Delta\mu_{\text{Pb}} + 2\Delta\mu_{\text{I}} &\leqslant \Delta H (\text{PbI}_2^{}) \, .
\end{split} \nonumber
\end{equation}
However, due to the unstable radical nature of neutral MA$\equiv\text{CH}_3^{}\text{NH}_3^{}$ (the reaction $\text{CH}_3^{}\text{NH}_2^{} + \frac{1}{2}\text{H}_2^{} \to \text{CH}_3^{}\text{NH}_3^{}$ is endothermic), we use the sum $(\mu_{\text{MeNH}_2^{}}^{\minuso} + \mu_{\text{H}}^{\minuso})$ instead of $\mu_{\text{MA}}^{\minuso}$\,, and similarly $(\Delta\mu_{\text{MeNH}_2^{}} + \Delta\mu_{\text{H}})$ instead of $\Delta\mu_{\text{MA}}$\,. The inequalities should then be rewritten as
\begin{equation}
\begin{split}
\Delta H (\text{MAPbI}_3^{}) &\leqslant \Delta\mu_{\text{MeNH}_2^{}} \leqslant 0 \, , \\
\Delta H (\text{MAPbI}_3^{}) &\leqslant \Delta\mu_{\text{H}} \leqslant 0 \, , \\
\Delta H (\text{MAPbI}_3^{}) &\leqslant \Delta\mu_{\text{Pb}} \leqslant 0 \, , \\
\Delta H (\text{MAPbI}_3^{}) &\leqslant 3 \Delta\mu_{\text{I}} \leqslant 0 \, ;
\end{split} \label{stab1}
\end{equation}
and
\begin{equation}
\begin{split}
\Delta H (\text{MAPbI}_3^{}) &\leqslant \Delta\mu_{\text{MeNH}_2^{}} + \Delta\mu_{\text{H}} \\ & + \Delta\mu_{\text{Pb}} + 3\Delta\mu_{\text{I}} \, , \\
\Delta\mu_{\text{MeNH}_2^{}} + \Delta\mu_{\text{H}} + \Delta\mu_{\text{I}} &\leqslant \Delta H (\text{MAI}) \, , \\
\Delta\mu_{\text{Pb}} + 2\Delta\mu_{\text{I}} &\leqslant \Delta H (\text{PbI}_2^{}) \, .
\end{split} \label{stab2}
\end{equation}

The inequalities in Eq.~(\ref{stab2}) can be rearranged as
\begin{equation}
\begin{split}
\Delta H (\text{MAPbI}_3^{}) - \Delta H (\text{MAI})
& \leqslant \Delta\mu_{\text{Pb}} + 2 \Delta\mu_{\text{I}} \\
& \leqslant \Delta H(\text{PbI}_2^{})\,, \\
\Delta H (\text{MAPbI}_3^{}) - \Delta H(\text{PbI}_2)
& \leqslant \Delta\mu_{\text{MeNH}_2^{}} + \Delta\mu_{\text{H}} + \Delta\mu_{\text{I}} \\
& \leqslant \Delta H (\text{MAI})\,. 
\end{split} \label{stab3}
\end{equation}

Inequalities in Eq.~(\ref{stab1}) define the domains of variables $\mu_{\text{MeNH}_2^{}}^{}$\,,  $\mu_{\text{H}}^{}$\,, $\mu_{\text{Pb}}^{}$\,, and $\mu_{\text{I}}^{}$, and the inequalities in Eq.~(\ref{stab3}) define the region for growth of ``stable-bulk MAPbI$_3^{}$'' in the phase diagram. $\mu_{\text{MeNH}_2^{}}^{\minuso}$\,,  $\mu_{\text{H}}^{\minuso}$\,, $\mu_{\text{Pb}}^{\minuso}$\,, and $\mu_{\text{I}}^{\minuso}$ can be calculated for the stable reference structures of MeNH$_2^{}$ (molecule), $\text{H}$ ($\text{H}_2^{}$ molecule), Pb ($P6_3^{}/mmc$), and I ($\text{I}_2^{}$ molecule) with DFT, respectively. Formation energies $\Delta H$ in Eq.~(\ref{stab3}) can be calculated with DFT, too.

Equations~(\ref{stab1}) and (\ref{stab3}) only serve to determine the bulk stability. For the stability of (clean and reconstructed) surface models, we need to solve Eq.~(\ref{grandpot.}) to obtain the SPDs. Note that the bulk and surface are not in isolation from each other. The final surface stability is determined by the intersection of the  SPD and the stable-bulk region. %The overlap of the stable-bulk region with the SPD can be regarded as the predictor of a viable bulk and surface together.

In principle, we need to plot the SPDs in four dimensions (4D) as the grand potential of each surface is a function of four variables ($\Delta \mu_{\text{MeNH}_2^{}}^{}$\,,  $\Delta \mu_{\text{H}}^{}$\,, $\Delta \mu_{\text{Pb}}^{}$\,, and $\Delta \mu_{\text{I}}^{}$). In practice, however, such a 4D diagram is hard to draw and visualize, and we use three two-dimensional (2D) slices instead: the $\Delta\mu_{\text{I}}^{}$/$\Delta\mu_{\text{MeNH}_2{}}^{}$ slice at $\Delta\mu_{\mathrm{Pb}} \!=\! \Delta\mu_{\mathrm{H}} \!=\! 0$\,, the $\Delta\mu_{\text{I}}^{}$/$\Delta\mu_{\text{Pb}}^{}$ slice at $\Delta\mu_{\mathrm{\mathrm{MeNH}_2^{}}} \!=\! \Delta\mu_{H} \!=\! 0$\,, and the
$\Delta\mu_{\text{I}}^{}$/$\Delta\mu_{\text{H}}^{}$ slice at $\Delta\mu_{\mathrm{Pb}} \!=\! \Delta\mu_{\mathrm{MeNH}_2^{}} \!=\! 0$\,.

\section{Results and Discussion}\label{results}

\subsection{Thermodynamic stability analysis of bulk and surface terminations}

The PBEsol-calculated formation energies of bulk MAPbI$_3^{}$\,, MAI, and PbI$_2^{}$ are $-4.82$, $-2.30$, and $-2.47$~eV, respectively. From Eq.~(\ref{stab3}), we can find the numerical values for thermodynamic growth limits of bulk MAPbI$_3^{}$ in its tetragonal phase:
\begin{equation*}
\begin{split}
-2.53~\text{eV} & \leqslant \Delta\mu_{\text{Pb}} + 2 \Delta\mu_{\text{I}} \leqslant -2.47~\text{eV}, \\
-2.35~\text{eV} & \leqslant \Delta\mu_{\text{MeNH}_2^{}} + \Delta\mu_{\text{H}} + \Delta\mu_{\text{I}} \leqslant -2.30~\text{eV}. \label{stab5}
\end{split}
\end{equation*}

The energy required for tetragonal MAPbI$_3^{}$ to decompose into MAI and PbI$_2^{}$, i.e., the difference between the left and the right values of either inequality, is as small as $0.06~\text{eV}$. Such a narrow stability region reflects the general instability of tetragonal MAPbI$_3^{}$.

\begin{figure}[!htp]
\includegraphics[width=\linewidth]{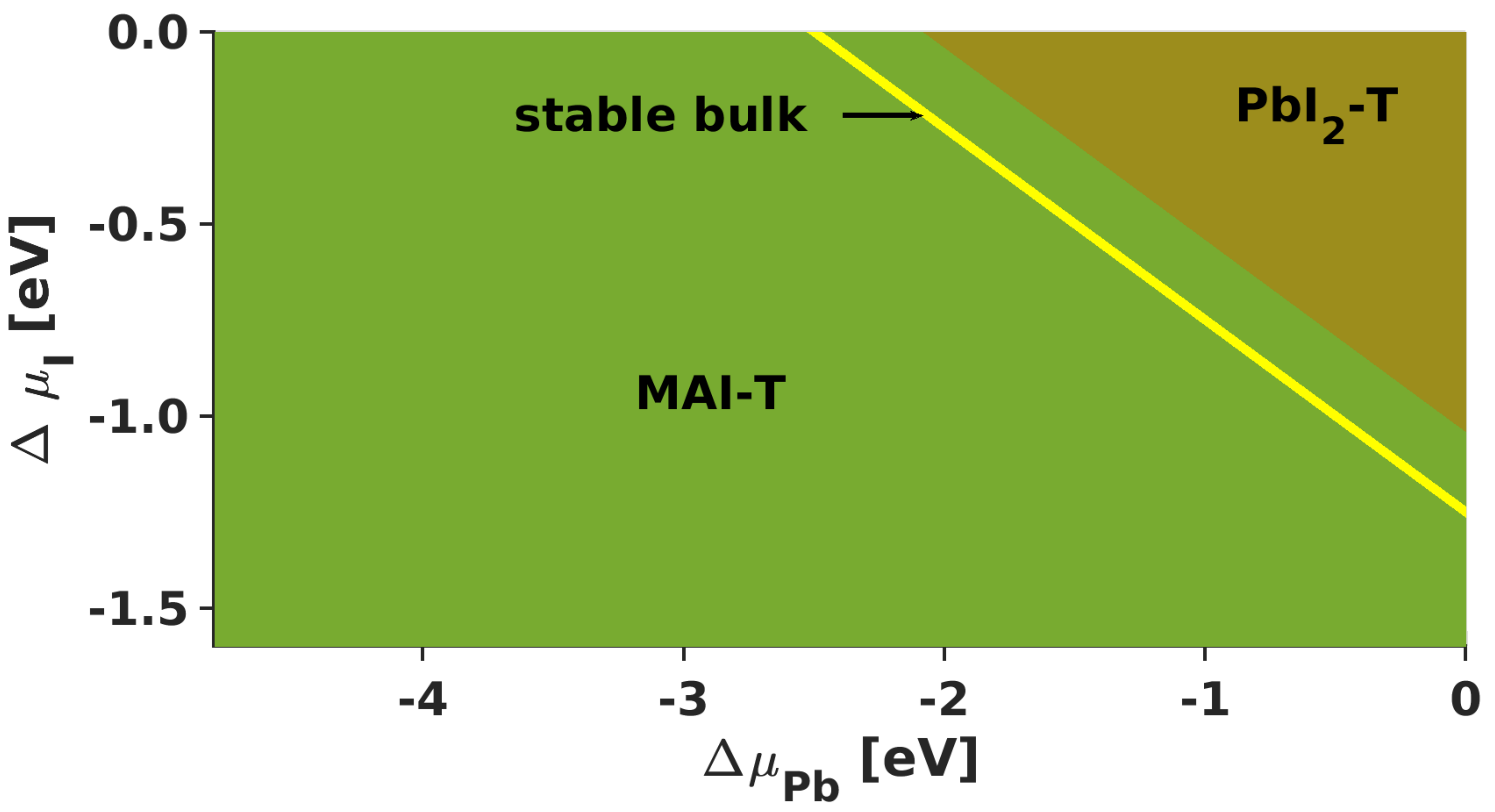}\vspace{-1.0em}
\caption{Thermodynamic growth limit for MAI-T and PbI$_2$-T surfaces in tetragonal MAPbI$_3$\,. The yellow shaded regions depict the thermodynamically stable range for the growth of bulk MAPbI$_3$\,.}\label{mai_pbi2}
\end{figure}

SPD analysis helps identify the stability of the two considered surface terminations. Figure~\ref{mai_pbi2} shows that, at $\Delta\mu_{\mathrm{MeNH}_2^{}} \!=\! \Delta\mu_{\mathrm{H}} \!=\! 0$, the MAI-T and PbI$_2$\,-T clean surfaces are stable in the Pb-poor and Pb-rich limits, respectively. We consider MAI-T the more stable surface since the region for stable MAI-T covers a wider range of $\Delta\mu_k$ ($k\!=\!$ Pb and I, as well as MeNH$_2^{}$ and H which are not shown here). Furthermore and quite importantly, the stable bulk region, shown by yellow shading in Figure~\ref{mai_pbi2}, intersects only the MAI-T surface. This finding agrees with previous theoretical results for $\text{MAPbI}_3^{}$ \cite{Yin14, He2020, Caddeo2020, Baikie2013, Mirzehmet2021, Geng15} that claimed the stability of MAI-T over PbI$_2$\,-T and is similar to the CsPbI$_3$ surface properties that we reported earlier \cite{Seidu2021}. Our discussions will therefore focus on MAI-T surfaces from here on. Data for PbI$_2$\,-T surfaces, including the relaxed surface-reconstructions and the SPDs, are given in the Supplementary Material (SM) .

\begin{figure*}[!ht]
\includegraphics[width=\linewidth]{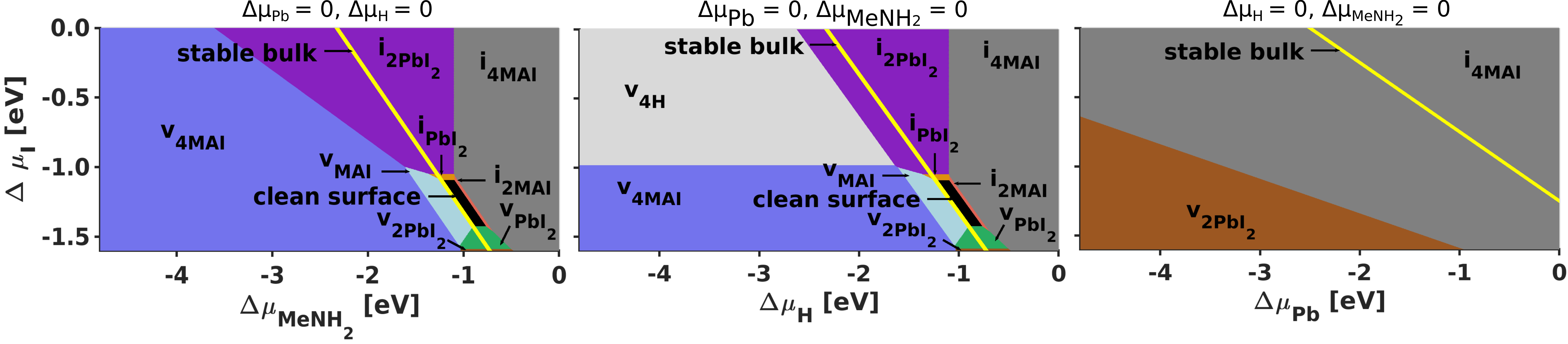}\vspace{-1.0em}
\caption{2D slices through the 4 dimensional surface phase diagrams of MAI-T surfaces of tetragonal MAPbI$_3$. The yellow regions in each panel marks the bulk stability region of MAPbI$_3$.  \label{spds}}
\end{figure*}

\subsection{Identification of stable reconstructions of MAI-T surfaces}

SPDs for the considered surface reconstructions of the MAI-T surfaces are shown in Figure~\ref{spds} (SPDs for the PbI$_2$\,-T counterparts are available in Fig.~S1 of SM). It is not surprising that the $\Delta\mu_{\text{I}}^{}$/$\Delta\mu_{\text{MeNH}_2{}}^{}$ and $\Delta\mu_{\text{I}}^{}$/$\Delta\mu_{\text{H}}^{}$ SPDs display similar features, as $\text{MeNH}_2^{}$ and $\text{H}$ are closely related to each other through the organic MA component of the material. In these two SPDs, which are given in the Pb-rich limit ($\Delta\mu_{\text{Pb}}^{}\!=\!0$), we observe the following stable surface structures: i$_{4\text{MAI}}^{}$ (in the $\text{MeNH}_2^{}$\,- and H-rich limit), v$_{4\text{MAI}}^{}$ (in the $\text{MeNH}_2^{}$\,- and H-poor limit), clean surface, v$_{\text{PbI}_2^{}}$\,, v$_{2\text{PbI}_2^{}}$\,, i$_{\text{PbI}_2^{}}$\,, i$_{2\text{PbI}_2^{}}$\,, i$_{2\text{MAI}}^{}$\,, and v$_{\text{MAI}}^{}$\,. The major difference in the appearance of these two phase diagrams lies with the v$_{\text{4H}}^{}$ surface, which is observed in the H-poor and I-rich limit. With our choice of 2D slices, this surface reconstruction appears in one quadrant of only one of these two 2D phase diagrams.

The MeNH$_2$\,- and H-rich (thus MA-rich) limit ($\Delta\mu_{\text{MeNH}_2^{}}^{}\!=\!\Delta\mu_{\text{H}}^{}\!=\!0$) creates a third 2D slice of the total phase diagram, shown on the right side of Figure~\ref{spds}. In this SPD, we find i$_{4\text{MAI}}$ and v$_{2\text{PbI}_2^{}}^{}$ to be stable. Except for v$_{\text{4H}^{}}$\,, all the observed stable reconstructions are valence-neutral, i.e., with addition or removal of MAI or PbI$_2$ units, net charges are not induced in the system, which is similar to what we previously found for the CsPbI$_3^{}$ surfaces \cite{Seidu2021}. We notice that in the $\text{MeNH}_2^{}$\,-, H-, and Pb-rich limit i$_{4\text{MAI}}^{}$ dominates over PbI$_2^{}$-derived reconstructions. That is, on the MAI termination layer at the MAPbI$_3^{}$ surface, the tendency for growing an extra MAI layer is greater than for growth of PbI$_2^{}$ units, which would eventually transform the system into PbI$_2^{}$\,-T. This finding again verifies that MAI-T is more stable.

We are particularly interested in the most relevant reconstructions, which we define as those regions in the SPDs that intersect the stable bulk region. It is these overlapping regions of bulk and surface stability that are viable standalone surfaces in the laboratory. These relevant models are the clean surface, v$_{\text{PbI}_2^{}}$\,, v$_{2\text{PbI}_2^{}}$\,, i$_{\text{PbI}_2^{}}$\,, and i$_{2\text{PbI}_2^{}}$ in the Pb-rich limit, and i$_{\text{4MAI}}$ in the MA-rich limit. Different from CsPbI$_3^{}$, for which we observe a relatively broad range in chemical potential for the clean surface, the range for its stability on MAI-T $(001)$ at $\Delta\mu_{\text{Pb}}\!=\!0$ is very narrow in terms of $\Delta\mu_{\text{MeNH}_2^{}}^{}$ and $\Delta\mu_{\text{H}}^{}$. In addition, it is only stable in I-deficient growth conditions.

\begin{figure*}[!htp]
\includegraphics[width=\linewidth]{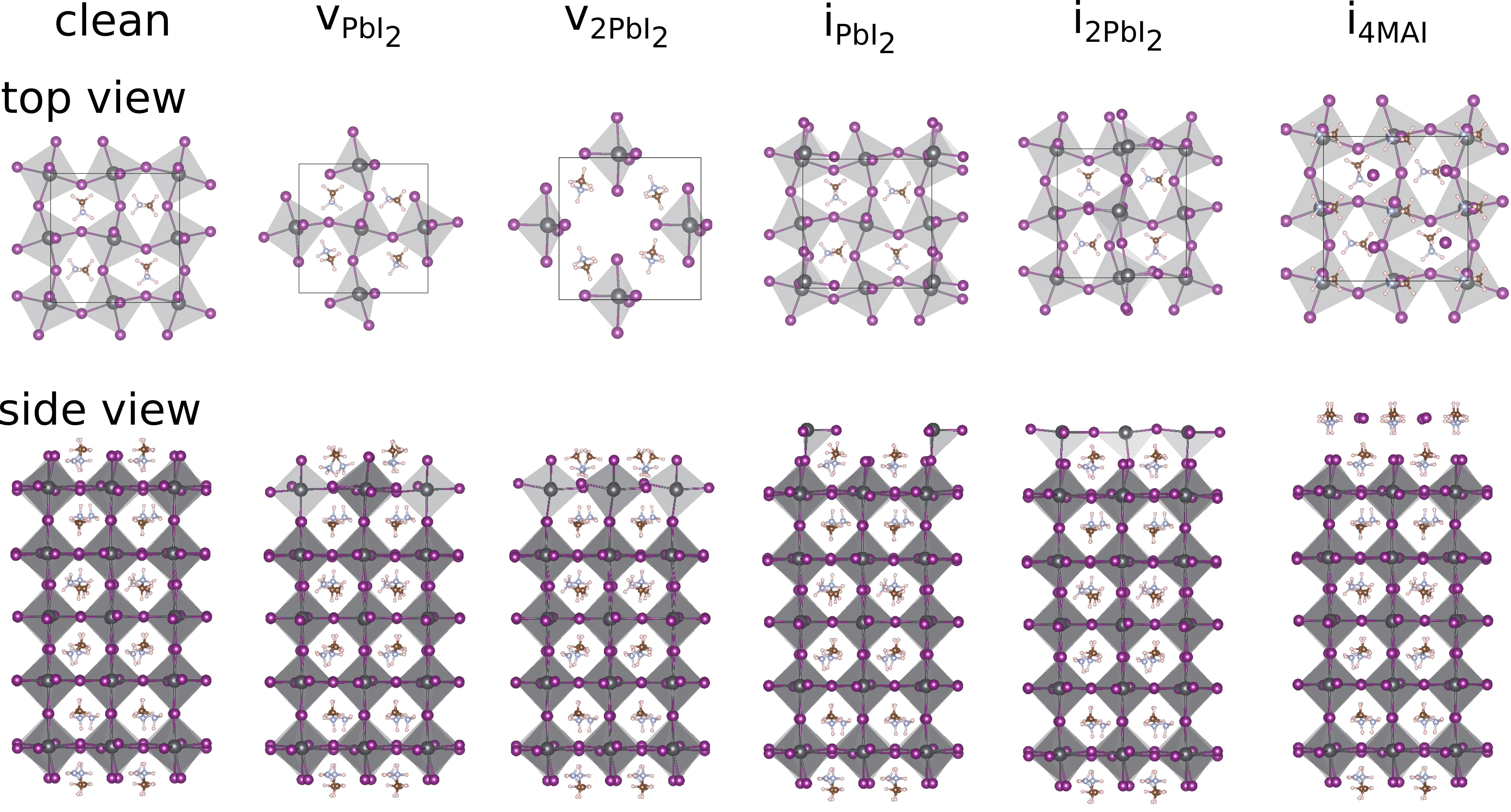}\vspace{-1.0em}
\caption{Atomic structures of the most relevant surface reconstruction models of tetragonal MAPbI$_3$\,. The atomic colour convention follows that of Fig.~\ref{bulk}. The light gray shades depict the octahedra in the topmost in the ``top view`` and sites with missing or add PbI$_2$-units in the ``side view``.   }\label{structures}
\end{figure*}

\begin{figure}[!ht]
\includegraphics[width=0.7\linewidth]{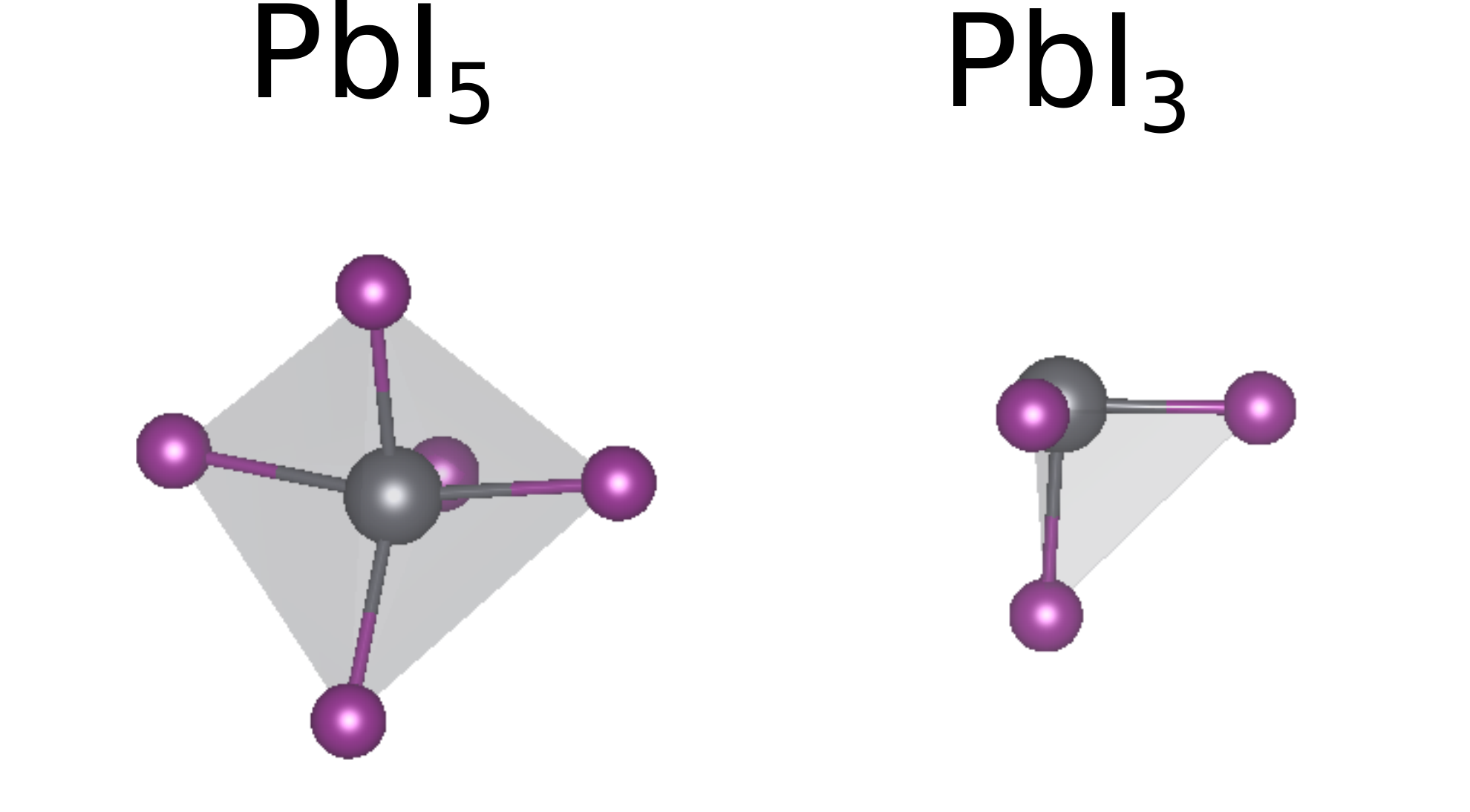}\vspace{-1.0em}
\caption{Surface polyhedra in reconstructed models with missing- and add-PbI$_2$\, units.}\label{poly}
\end{figure}

The optimized geometries of the most relevant surface models are given in Figure~\ref{structures} (relevant reconstruction models of PbI$_2$\,-T are presented in Fig.~S2 of SM). Because the surface atomic structure varies mainly to accommodate the absence or addition of atoms, our discussion of geometric rearrangement will be with reference to the clean surface in the following. We observe that PbI$_2^{}$ removal causes noticeable atomic structure changes in the reconstructed surfaces. The topmost PbI$_2^{}$ layer of v$_{\text{PbI}_2^{}}$ displays $(\text{PbI}_6^{})_2^{}(\text{PbI}_5^{})$ polyhedra, while in v$_{2\text{PbI}_2^{}}$ there are two isolated PbI$_5^{}$ polyhedra (Fig.~\ref{poly}). Interestingly, no migration of surface I anions occurs in v$_{2\text{PbI}_2^{}}$, which is the main characteristic of the equivalent removal on $\upalpha$-CsPbI$_3^{}$ \cite{Seidu2021}. This is very likely due to the different A-site cations: the hydrogen bonding between the ammonium group and the surface I anions would stabilize the latter, so that the surface Pb--I units are relatively regularly distributed.

In i$_{\text{PbI}_2^{}}^{}$ and i$_{2\text{PbI}_2^{}}^{}$\,, each added PbI$_2^{}$ unit is linked to a surface I atom via Pb--I bonding, giving rise to a PbI$_3^{}$ tetrahedron that contains one Pb and three I atoms as its vertices (Figure~\ref{poly}). Notably, the i$_{2\text{PbI}_2^{}}^{}$ reconstruction shows characteristic PbI$_5^{}$PbI$_3^{}$ polyhedra (Fig.~\ref{poly}), as previously reported by Haruyama et~al. \cite{Haruyama14}. This asymmetric distribution of surface I atoms results from the removal of two linearly-aligned PbI$_2^{}$ units, which is very different to the same surface reconstruction of $\upalpha$-CsPbI$_3^{}$ where the diagonal mode is more stable.

Finally, we find a relatively regular alignment of the added MAI units, forming a uniform sheet on the MAI-T surface in i$_{4\text{MAI}}$\,. The C--N bonds of all added MA$^+$ cations point towards the surface to form hydrogen bonds with the topmost I anions. The average shortest $\text{H}(\text{N})\cdots\text{I}$ distance is $2.68~\text{\AA}$, a typical value for hydrogen bonding \cite{li,li2018b}. The fact that extra MAI units can readily grow above the already existing MAI surface-termination layer, as also indicated by the SPDs in Fig.~\ref{spds}, makes it very likely that MAI multilayers can grow on  MAPbI$_3^{}$ surfaces in MAI-rich situations. This would be detrimental to device performance since MAI is very poor in transporting charge carriers.

\subsection{Electronic properties of bulk and most relevant reconstructed \texorpdfstring{MAPbI$_3$}{} surface models}\label{elec_prop}

In this section, we focus on the electronic properties of bulk MAPbI$_3$\, and the most relevant reconstructed MAI-T surface models. The band structures of the most relevant PbI$_2$\,-T surface reconstructions are presented in Fig.~S3 of SM.

\subsubsection{Electronic properties of the bulk and the clean MAI-T surface}

\begin{figure*}[!ht]
\includegraphics[width=\linewidth]{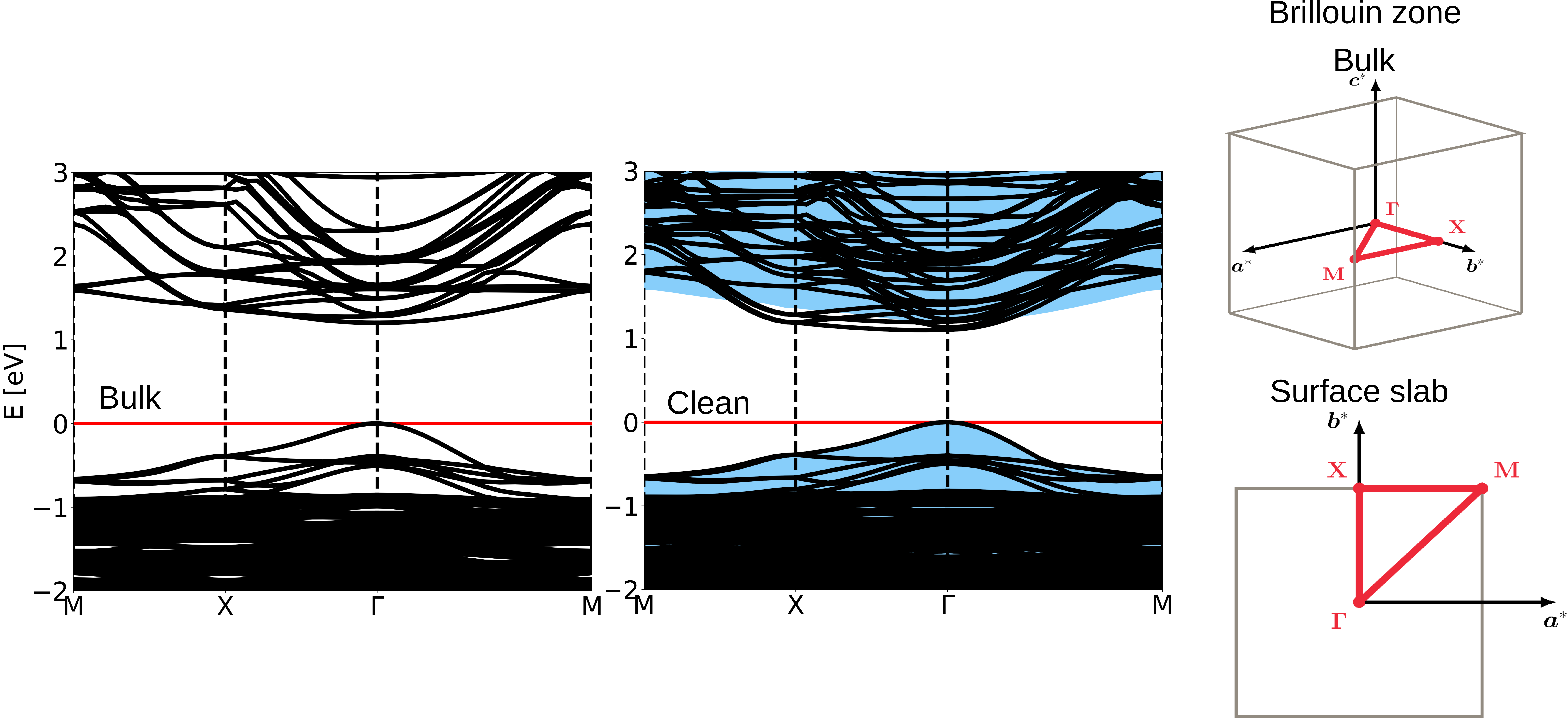}\vspace{-1.0em}
\caption{Band structures of bulk and clean MAI-T surface of tetragonal MAPbI$_3$\,. Both bulk and surface band structures are calculated with a $2\times2$ in-plane supercell to share a common Brillouin zone and $k$-point path (far right). VBM is set to 0 as marked by the red horizontal line. In the surface band structure plots, the projected bulk band structure is shown as blue shading.}\label{bands}
\end{figure*}

Figure~\ref{bands} depicts the band structures of the bulk and the pristine MAI-T surface of MAPbI$_3$\,. For the bulk, we adopt the high-symmetry $k$-point path of a simple-cubic lattice in the $2\times2\times2$ supercell model for simplicity. Our plots show the band structure along M--X--$\Gamma$--M with $\text{M}\!=\!\big(\frac{1}{2},\frac{1}{2},0\big)$, $\text{X}\!=\!\big(0,\frac{1}{2},0\big)$, and $\Gamma\!=\!(0,0,0)$, i.e., within the $a^{\ast}b^{\ast}$ plane of the Brillouin zone (identical to the $ab\!=\!(001)$ plane in real space in our cases).

\begin{figure}[!ht]
\includegraphics[width=\linewidth]{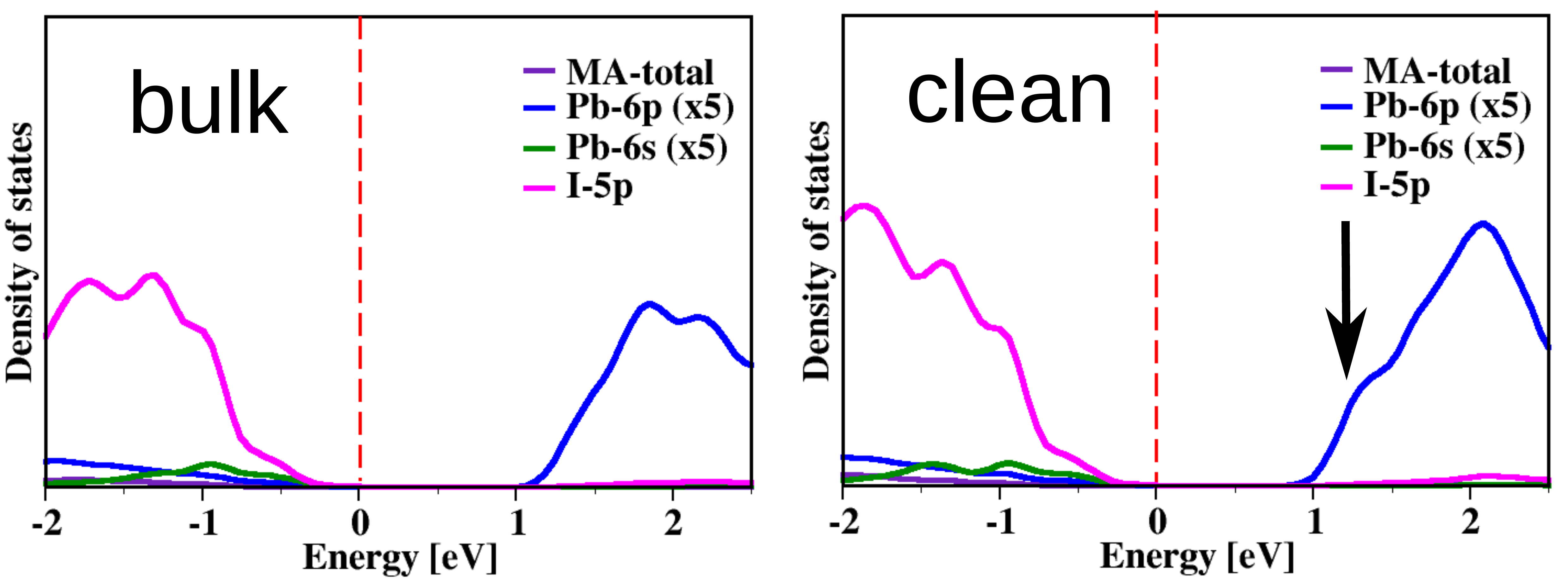}\vspace{-1.0em}
\caption{Density of states of the bulk and clean MAI-T surface of tetragonal MAPb$_3$\,, showing the contributions of different atomic species. We scaled the Pb density of states by a factor 5 (depicted by *5) to make it more visible The arrow is indicative of the surface states at the edge of the CB in the clean surface.}\label{dos}
\end{figure}

The bulk band structure exhibits a direct band gap at the $\Gamma$ point. The element projected density of states (PDOS)  in Fig.~\ref{dos} reveals that the valence band (VB) is dominated by I-5$p$ orbitals. The VB maximum (VBM) exhibits a noticeable contribution from Pb-6$s$ which gives rise to the well-known antibonding character and thus introduces the noticeable band dispersion at $\Gamma$ (Fig.~\ref{bands}). The conduction-band minimum (CBM) consists mainly of Pb-6$p$ orbitals. The MA cation shows no significant contributions at the band edges.

Next, we investigate, if our MAI-T surface model introduces surface or mid-gap states. The bands of the clean MAI-T surface are shown in the middle panel of Fig.~\ref{bands}. Both the VB and CB edges of the surface nearly align with the bulk bands at the M point. The band gap of the surface at M is slightly larger than the bulk, which could be a quantum confinement artefact of the slab model, but the bands themselves agree. At $\Gamma$ and X, however, the CB of the clean surface extends below the CBM of the bulk. At these points, the shapes of the bulk and surface bands at $\Gamma$ are different.

\begin{figure}[!ht]
\includegraphics[width=\linewidth]{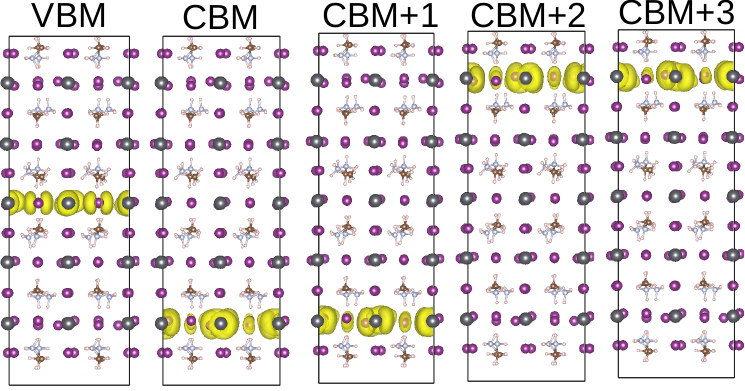}\vspace{-1.0em}
\caption{Charge distributions of the valence band maximum (VBM) and conduction band minimum (CBM) up to CBM+3 at $\Gamma$ point for the clean surface.}\label{clean_cbm}
\end{figure}

To understand the nature of these states, we first analyse the PDOS in Fig.~\ref{dos}. The PDOS verifies that the apparent band gap of the clean surface is $\sim\!0.2$ eV less than the bulk, as already indicated by the band structure. We plot the charge densities of the lowest four CB states of our surface model in Fig.~\ref{clean_cbm}. We find that these states are surface states that  come in nearly degenerate pairs, with the partners of each pair on opposite sides of the slab. The slight degeneracy lift in each pair is caused by a small relaxation induced structural asymmetry in the two bulk halves that make up our surface slab model (see Section~\ref{sec:bulkstructure}). The characters of these band edge wave functions are the same as the ones we observed for the $\upalpha$ phase of CsPbI$_3$ \cite{Seidu2021}. 

The DOS of the clean surface in Fig.~\ref{dos} is consistent with the interpretation of conduction band derived surface states. The right panel exhibits small bumps as shown by the arrow in Fig.~\ref{dos} (right panel) in the CB at approximately 1.25  and 2.10 eV compared to the bulk, indicating the rearrangement of bands in the slab model. 

The behaviour of the VBM is a little different. It is pinned to the domain wall at the mid-plane of the slab model. This quasi-two-dimensional state is still dispersive and its band very closely matches that of the bulk, which provides good evidence for the quality of our slab model. The pinning of the state to the mid-plane is reasonable. The dipoles on either side of the domain wall point away from the mid-plane, raising the energy of an electron residing on the mid-plane and positioning it as the maximum of the occupied states (VBM). Again, the lack of artefacts in the slab electronic structure and good agreement with the bulk support the quality of the model, even with the pinning of the VBM to the domain wall.

\subsubsection{Electronic properties of the most relevant MAI-T surface models}

\begin{figure*}[!htp]
\includegraphics[width=\linewidth]{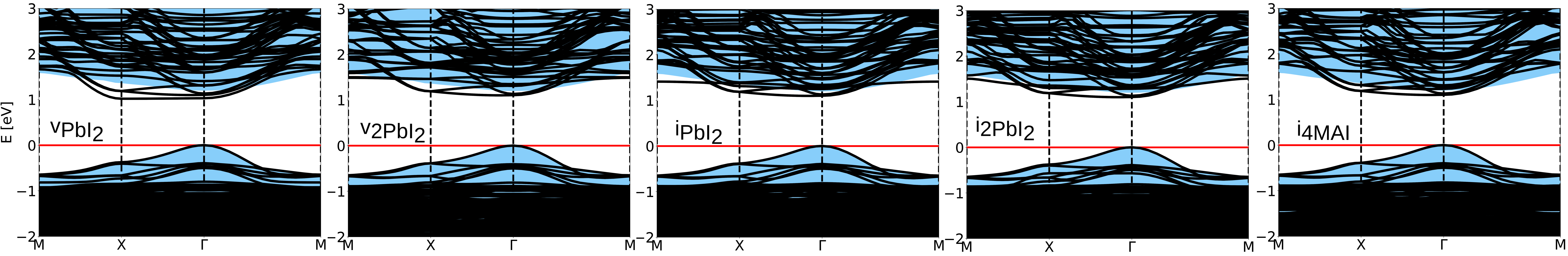}\vspace{-1.0em}
\caption{Band structures of the most relevant surface reconstruction models in tetragonal MAPbI$_3$\,. The bulk-projected band structure is depicted by the blue shading.}\label{band_rel}
\end{figure*}

\begin{figure}[!htp]
\includegraphics[width=\linewidth]{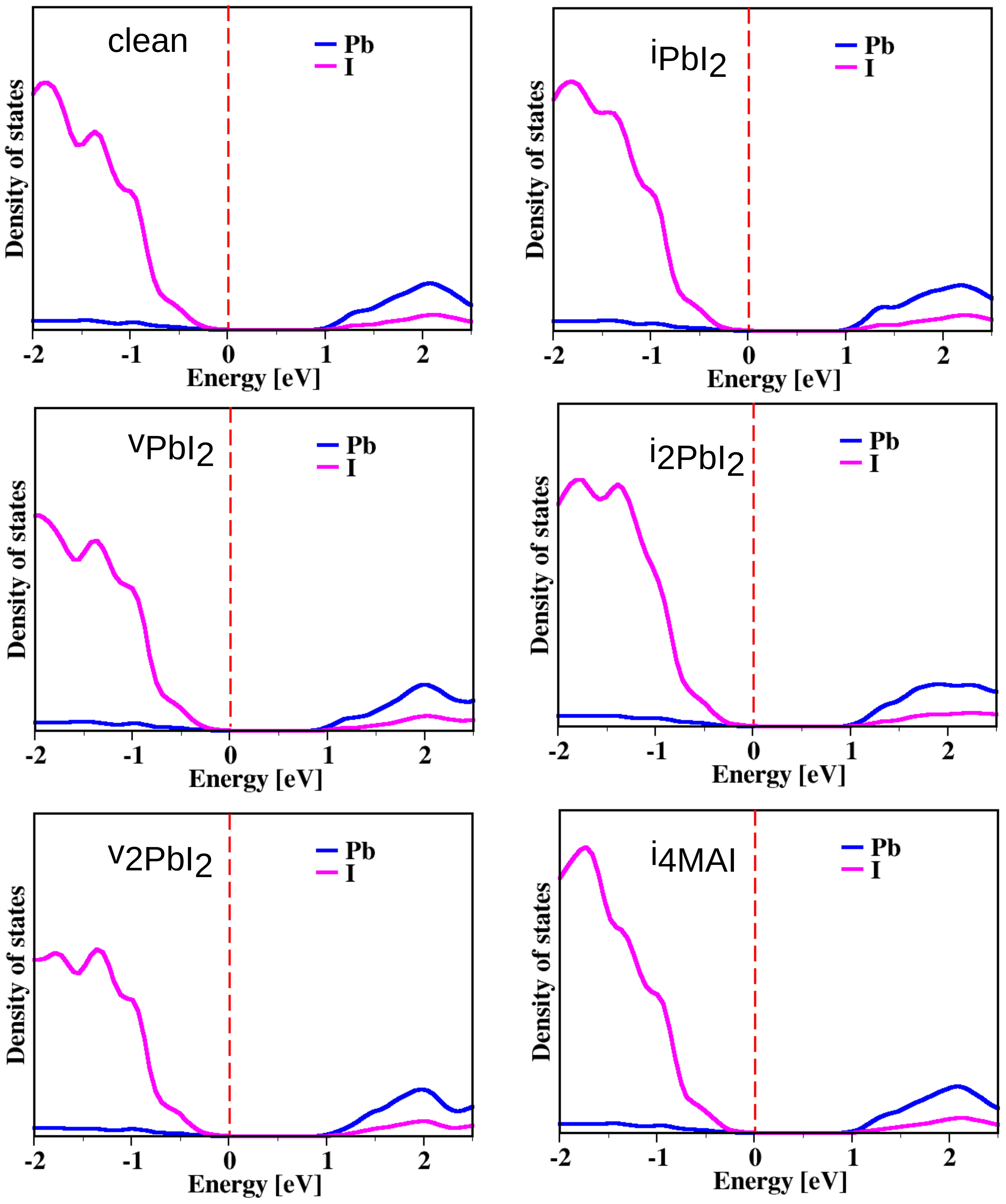}\vspace{-1.0em}
\caption{Density of states of the most relevant surface reconstruction models in MAI-T tetragonal MAPbI$_3$\,. The VBM is set to zero and shown as a red dashed line. }\label{pdos_rel}
\end{figure}

The band structure and PDOS of the most relevant surface models observed in Fig.~\ref{spds}, (v$_{\mathrm{PbI}_2^{}}$\,, v$_{2\mathrm{PbI}_2^{}}$\,, i$_{\mathrm{PbI}_2^{}}$\,, i$_{2\mathrm{PbI}_2^{}}$\,, and i$_{4\mathrm{MAI}}$\,), are shown in Figures~\ref{band_rel} and \ref{pdos_rel}, respectively. Similar to Fig.~\ref{bands}, the bulk band structure is included as background for comparison in Fig.~\ref{band_rel}. We find that the band structures of two of the most relevant reconstructed surface models, v$_{\mathrm{PbI}_2^{}}$ and i$_{4\text{MAI}}$\,, resemble the band structure of the clean MAI-T surface shown in Fig.~\ref{bands}.

\begin{figure*}[!ht]
\includegraphics[width=0.7\linewidth]{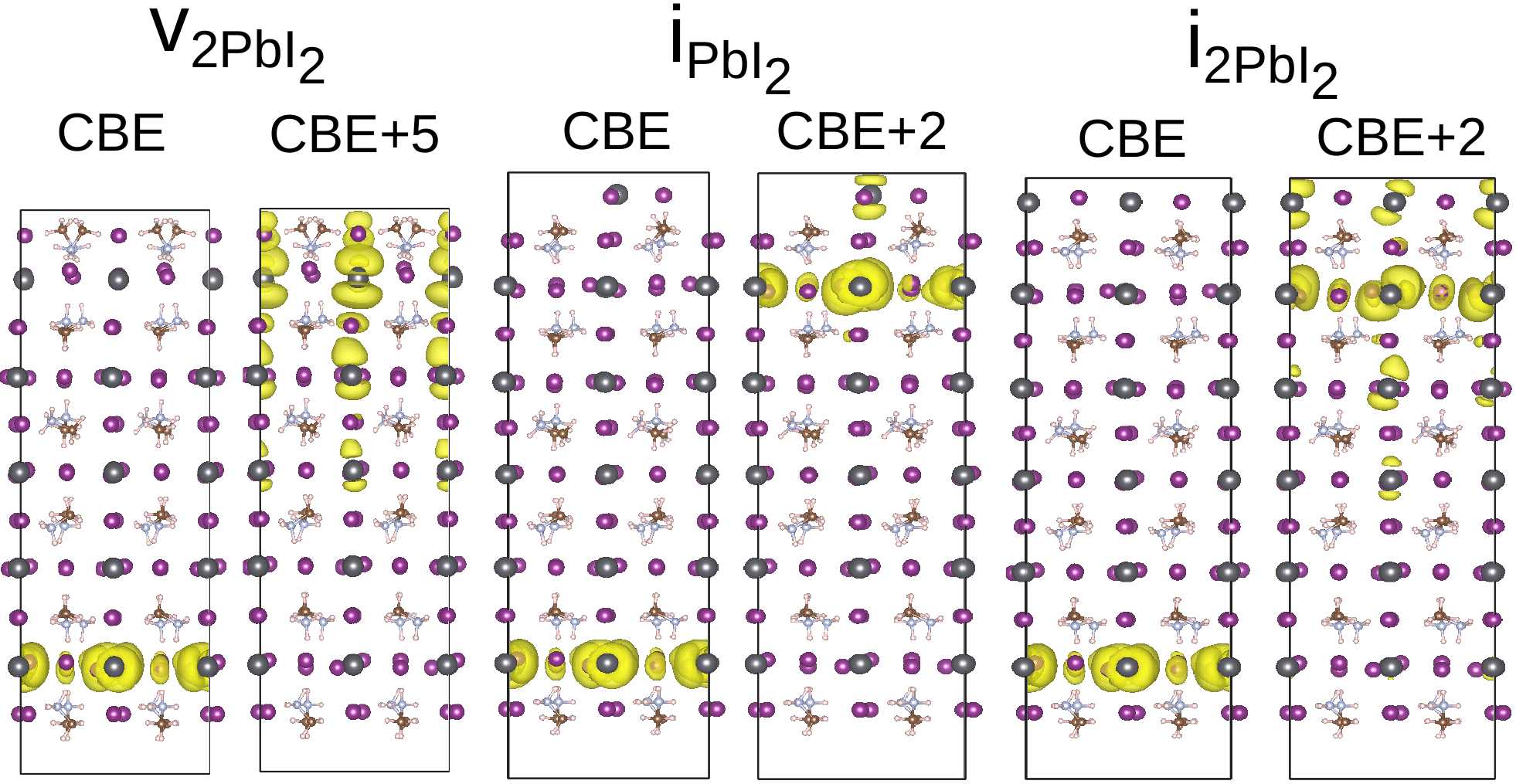}\vspace{-1.0em}
\caption{Charge distribution for v$_{2\text{PbI}_2^{}}$, i$_{\text{PbI}_2^{}}$ and i$_{2\text{PbI}_2^{}}$ of the conduction band edges (CBE) at M-point. We also show CBE+5 v$_{2\text{PbI}_2^{}}$ as well as CBE+2 for i$_{\text{PbI}_2^{}}$ and i$_{2\text{PbI}_2^{}}$.}\label{ipbi2_charge}
\end{figure*}

For the others, there are flat bands near and below the bulk CB edges (CBEs), which are more pronounced at the M-point. They are most visible in i$_{\mathrm{PbI}_2^{}}$. 
In the i$_{\mathrm{PbI}_2^{}}$ and i$_{2\mathrm{PbI}_2^{}}$ surfaces, we observe increased intensities within the CB of the PDOS at $\sim\!1.6$ and $\sim\!1.9$ eV, which corresponds to the flat bands at M-point in Fig.~\ref{band_rel}. These small peaks come from Pb states, suggesting that the flat bands are indeed due to the added PbI$_2$ units. To confirm this, we plot the charge distribution of CBE at M for these two reconstructions in Figure~\ref{ipbi2_charge}. 

Even though the CBE at M in these two reconstructions and in v$_{2\text{PbI}_2^{}}^{}$ belongs to a band which is flat across the entire Brillouin zone, the wave functions of these states (Fig.~\ref{ipbi2_charge})  still resemble the surface states of the pristine slab, especially for v$_{2\text{PbI}_2^{}}^{}$ and i$_{2\text{PbI}_2^{}}^{}$. This is somewhat surprising since the CBE states of the clean-surface model belong to a dispersive band. The CBE states of the reconstructed surfaces in Fig.~\ref{ipbi2_charge} still come in nearly degenerate pairs with each partner appearing on opposite sides of the slab, as for the clean surface. However, we can ignore the state at the bottom of the slab, since it corresponds to the clean and not the reconstructed surface. The states in i$_{\text{PbI}_2^{}}^{}$ and i$_{2\text{PbI}_2^{}}^{}$ at the top of the slab, on the reconstructed surface side, have considerable weight on the added PbI$_2$ units. This wave-function localization explains the flat character of the band.

\section{Conclusion}\label{conclusion}
In summary, we have investigated the  stability and electronic structure of MAPbI$_3$ surfaces in the tetragonal phase from first principles. To circumvent the polar catastrophe in our supercell calculations, we build a slab geometry from two MAPbI$_3$ bulk segments with opposite polarity, effectively introducing a domain wall in the middle of the slab. Our surface science study reveals that the methylammonium-iodine (MAI) termination is more stable than the PbI$_2$ termination. We further observe that the removal or addition of polar units that induce zero net charge in the system lead to more stable surface reconstructions. MAI-terminated surfaces introduce conduction-band derived surface states near the conduction band edge, which result in a surface band gap that is slightly smaller than the bulk band gap. The stable reconstructions do not introduce further surface states in the band gap, which bodes well for the transport properties across interfaces with these reconstructions. Our study opens up future work on surface adsorbates, defects and interfaces.

\section{Supplementary Material}
See Supplementary Material for  surface phase diagrams of PbI$_2$-terminated models, crystal  and electronic band structures of the most relevant reconstructed surfaces of PbI$_2$-T models .

\section{Authors contributions}
All authors contributed equally in this work.

\section*{Acknowledgments}
We acknowledge the computing resources from the CSC-IT Center for Science, the Aalto Science-IT project, and Xi’an Jiaotong University’s HPC Platform. We further acknowledge funding  from the  V\"ais\"al\"a Foundation and the Academy of Finland through its Key Project Funding scheme (305632) and postdoctoral grant no.~316347. 

\section{Conflict of interest}
The authors have no conflicts to disclose.

\section{Aip publishing data sharing policy} 
The data that supports the findings of this study will be openly available in Novel Materials Discovery (NOMAD) repository at \textcolor{red}{\cite{Note-NOMAD}}.

\newpage
\bibliography{aipsamp}% Produces the bibliography via BibTeX.

\end{document}